\documentstyle[12pt]{article}

\begin{document}

\rightline{CERN-TH.7556/95}

\begin{center}

{\large\bf Higgs Boson Production in
{\boldmath $e^+ e^- \to \mu^+ \mu^- b \bar b$} }
\vskip 4pt
{\large }

\end{center}

\noindent
Guido MONTAGNA$^a$, Oreste NICROSINI$^{b,}\footnote{\footnotesize
On leave from INFN, Sezione di Pavia, Italy.}$
and Fulvio PICCININI$^c$ \\

\noindent
$^a$ INFN, Sezione di Pavia,  Italy  \\
$^b$ CERN, TH Division, Geneva, Switzerland \\
$^c$ Dipartimento di Fisica Nucleare e Teorica,
Universit\`a di Pavia, and \\ INFN, Sezione di Pavia, Italy

\begin{abstract}

{\small  The production of the Standard Model
Higgs boson in the four-fermion
reaction $e^+ e^- \to \mu^+ \mu^- b \bar b$ is
studied. The complete tree-level matrix element, including
 signal and backgrounds in the standard electroweak theory, is computed and
initial state radiation is taken into account in the leading-log
approximation. A Monte Carlo event generator has been built and numerical
results for some  distributions of experimental interest for the search of the
Higgs particle at future electron-positron colliders are shown, compared
with those existing in the literature and commented.}

\end{abstract}

\vskip 24pt
\begin{center}
Submitted to Physics Letters B
\end{center}

\vskip 48pt \noindent
E-mail: \\
montagna@pv.infn.it \\
nicrosini@vxcern.cern.ch  \\
piccinini@pv.infn.it \\

\vfil
\leftline{CERN-TH.7556/95}
\leftline{January 1995}

\eject

After the experimental confirmations of the standard theory of the electroweak
interaction provided by the precision tests at
LEP/SLC~\cite{dsc} and by the evidence of
top quark production at the TEVATRON~\cite{prlcdf},
the search for the Standard Model Higgs
boson becomes a primary task of the experiments planned at future
electron-positron and hadron colliders. For the case of a light
or intermediate Higgs particle ($M_H \leq 2 M_W$),
the Bjorken bremsstrahlung process $e^+ e^- \to H Z$
is known to be the main production mechanism at $e^+ e^-$
colliders up to 0.5~TeV~\cite{djou}. Detailed studies about the production of
the Higgs boson in $e^+ e^-$ collisions
are given in Refs.~\cite{zal,cin}. From these investigations
it emerges that the specific four-fermion reaction
\begin{equation}
e^+ e^- \to H Z \to \mu^+ \mu^- b \bar b
\label{eq:react}
\end{equation}
provides the cleanest event signature coupled with
``small'' backgrounds.

A complete
tree-level calculation of the four-fermion process~(\ref{eq:react}),
including all possible diagrams and their interferences,
 has been performed in two recent papers~\cite{russi,blr}.
On the one hand, in Ref.~\cite{russi}
 a Monte Carlo approach has been adopted but the effect of the
initial state QED corrections, large and hence
 necessary for the data analysis, has been neglected; on the other hand,
 initial state radiation has been included in~\cite{blr}, but
 according to a semi-analytical procedure
 which is limited from the point of view of event simulation purposes.
The aim of this letter is to present a Monte Carlo event generator
for the process $e^+ e^-  \to \mu^+ \mu^- b \bar b$
based on the calculation of the full set of diagrams contributing
in the standard electroweak theory to the assigned four-fermion final state
and on the inclusion of initial state QED
corrections in the leading-log approximation. We will only sketch the basic
theoretical ingredients, putting more emphasis on the phenomenological
aspects of the study and on the potentials of a ``specialized''
event generator
for future searches of the Higgs boson at LEP2 and NLC energies.

The kinematics of the $2 \to 4$ process at the Born level has been treated
according to the general recipe for a four-fermion final state
described in~\cite{noiww}. Essentially,
seven independent dimensionless quantities
are introduced and chosen, together with a trivial overall azimuthal
angle, as independent variables from which the full four-momenta
of the outgoing particles are reconstructed in the laboratory
frame by explicit solution of the kinematics.

The inclusion of initial state radiation is simply achieved
by dressing the incoming
leptons with QED structure functions~\cite{sf}.
They allow to include, by virtue of factorization theorems, the long-distance
universal contributions in soft and/or collinear approximation.

Taking into account the initial state radiation,
 the total cross section can be expressed
as a nine-fold convolution of the form

\begin{eqnarray}
\sigma (s) = \sum_i \int d x_1 \, d x_2 \, D(x_1,s) D(x_2,s)
d [PS] {{d \sigma} \over {d [PS]}} {{w_i} \over {W}} ,
\end{eqnarray}

\noindent
where $d [PS]$ denotes the volume element in the
7-dimensional phase space connected to the independent variables of the
hard scattering reaction in the c.m. frame. The scale of the
hard scattering cross section is the reduced c.m. energy
${\hat s} = x_1 x_2 s$ which is the main effect of the initial state radiation
together with the boost of the c.m. frame with respect to the laboratory one,
which is also taken into account by properly rescaling the independent
variables of the hard kernel cross section in the c.m. frame.

In the kernel cross
section the complete set of tree-level diagrams contributing in
the Standard Model to $e^+ e^- \to \mu^+ \mu^- b \bar b$ is included.
Twenty-five diagrams have to be computed and
they can be classified as follows:
\begin{itemize}

\item the signal diagram $e^+ e^- \to HZ$ followed by the decays
$H \to b \bar b$ and $Z \to \mu^+ \mu^-$;

\item the eight $t-$channel diagrams relative to
$ZZ, \gamma\gamma, Z\gamma$ exchange;

\item the sixteen $s-$channel diagrams corresponding to
the bremsstrahlung of a neutral vector boson from the fermionic final states
 (radiative processes).

\end{itemize}

Each amplitude has been computed by means of
{\tt SCHOONSCHIP}~\cite{schoon} within the helicity amplitude
formalism~\cite{hel} following the
general strategy described in~\cite{noiww}. This allows
to express the kernel cross sections essentially as the sum of products of
resonating propagators with functions of the dimensionless variables
originating from the calculation of the traces of the $\gamma$
matrices' strings. Mass corrections have been neglected in the kinematics
 and in the calculation of the amplitudes.
These contributions introduce terms of the order of $m_b^2 / s$ which
at LEP2 energies amount to about 0.1\% and hence they can be
 neglected as long as the ratio signal / background is larger than
some per cent.

To take care of the peaking behaviour of the integrand the Monte Carlo
event generation is treated according to the importance sampling
tech\-ni\-que\cite{james}.
The latter is applied, by means of proper changes of variables,
to the exponentiated infrared sensitive part of the structure functions,
 to the different propagators' ``singularities'' (particularly to the
Higgs propagator which resembles for its very narrow width
a delta-like function) and finally to
the jacobian peaks appearing in the phase space factor. In particular, the
weights $w_i$ and $W = \sum_i w_i$ are introduced to obtain a proper sampling
of the double propagator ``singularities'' of the signal ($ZH$) and the
backgrounds ($ZZ$, $Z \gamma$ and $\gamma \gamma$ in all possible
combinations).

Let us show and comment some results obtained
with our event generator based on the random number
generator {\tt RANLUX}~\cite{rlux}. The input parameters used are:
$M_{Z} = 91.1887\,$GeV, $\Gamma_{Z}
= 2.4974\,$GeV, $M_W = 80.22\,$~GeV,
$m_b = 4.7\,$~GeV~\cite{pdg}. For the Higgs boson width
 the tree-level expression is adopted.

In Fig.~1a the total cross section in Born approximation
is compared with the initial state QED-corrected one
as a function of the c.m. energy and for three different values
of the Higgs mass. The dashed, dash-dotted and closely dotted lines
 represent the lowest-order results for $M_H=65, 100, 140\,$~GeV, respectively;
the three solid lines refer to the above Higgs mass values
and include the effect of the leading-log initial state QED corrections.
As a reference curve, the QED corrected total cross section obtained
by excluding the Higgs boson contribution is also shown (sparsely dotted line).
Cuts on the invariant masses of the $\mu^+ \mu^-$ and $b \bar b$ pairs of
$M_{\mu^+ \mu^-} \geq 12$~GeV and $M_{b \bar b} \geq 12$~GeV (typically
used in other numerical studies~\cite{russi,blr})
are applied in order to reduce a large fraction of
the $\gamma \gamma$ background.
 The values obtained for the cross sections with our event generator
 have been compared in detail (i.e. for different Higgs masses and c.m.
 energies) with the independent  results of
Refs.~\cite{russi,blr} and found to be in good agreement.
It should be noted that in the LEP2 energy range the size of the total
cross section of the $\mu^+ \mu^- b \bar b$ reaction can give by itself
evidence of Higgs boson production in the ``light mass'' interval
$65$~GeV~$\leq M_H \leq 100$~GeV (see also Fig.~1b),
whereas for higher Higgs mass values a c.m. energy
larger than 200~GeV is necessarily required to distinguish the signal
from the background. This feature, already pointed out in Ref.~\cite{russi}
for the lowest-order cross section, is not largely affected
by initial state bremsstrahlung which essentially lowers the peak
cross section of about 10\% and originates a moderate radiative tail
at high energies ($\sqrt s \geq$~300~GeV ),
as generally occurs for other ``double resonant'' processes
($WW, ZZ$) above threshold. Moreover, the two-peak shape for $M_H= 65$~GeV and
$M_H= 140$~GeV is made more flat by effect of the initial state radiation,
which however does not cancel the two-bump structure. Their appearing
is given, with very good approximation, by the incoherent sum of the
signal and backgrounds diagrams. In agreement with Ref.~\cite{russi},
we checked that the interference terms between diagrams
with different intermediate bosons and also $s-$channel radiative processes
 give a small contribution.

Figure~1b shows the ratio $\sigma_s / \sigma_f$ as a function of
the Higgs mass and for three different energies of interest
at LEP2. $\sigma_s$ denotes the signal cross section
and $\sigma_f$ stands for the full four-fermion cross section including
signal and backgrounds. The effect of the initial state
radiation is included in this plot. As already
singled out by Fig.~1a, for the case
of a light Higgs mass ranging between, say 65 GeV and 80 GeV, the ratio
$\sigma_s / \sigma_f$ turns out to be more favourable for ``low'' centre
of mass energies; on the contrary ``high''
centre of mass energies maintain the ratio more flat and sensitive
to a Higgs mass up to, say, 100-110 GeV.

In Figs.~2-5 we show some
realistic distributions of experimental interest
for the Higgs search at LEP2.
These results have been obtained processing the $n$-tuple
created by the event generator at $\sqrt{s} = 205$~GeV. The full set
of tree-level diagrams and the contribution of
the initial state radiation have been taken into account in the
simulation. A sample
of $10^4$ four-fermion events has been generated in order to study with
small statistical error the behaviour of the distributions when varying the
Higgs mass. Although different cross sections correspond to
different Higgs mass values, the same number of events has been considered
 in order to study the pattern of evolution of the shapes while varying
the Higgs mass.
The $b \bar b$ invariant mass distribution (Fig.~2),
the $b-$quark scattering angle distribution (Fig.~3)
and the $b \bar b$ relative angle distribution (Fig.~4) turn out to be
useful observables to detect a Higgs signal and, if found, extract
information on the quantum numbers assignment.
{}From Fig.~2 it can be seen that the $b \bar b$ system invariant mass
distribution exhibits two clear spikes positioned at $M_H$ and $M_Z$, as
long as $M_H <$ say 110~GeV, the spike at $M_H$ resembling a delta-like
function. The rise at low invariant masses is due to $\gamma \to b \bar b$ in
$\gamma \gamma$ and $\gamma Z$ diagrams.
{}From Fig.~3 it can be seen that the $b-$quark angular distribution
shows, for the case of a
light mass Higgs where the signal
dominates the $\mu^+ \mu^- b \bar b$
cross section, a clearly isotropic spin-zero behaviour
 gradually disappearing as increasing the Higgs mass.
The $b \bar b$ relative angle distribution represented in Fig.~4
shows a two-peak structure. The peak positioned at
lower $\cos \vartheta_{b \bar b}$
reflects the presence of $Z Z$ diagrams, while the (much smaller) one at
$\cos \vartheta_{b \bar b} \approx 1$ comes
from $\gamma \to b \bar b$ in the $\gamma \gamma$ and $\gamma Z$
contributions.
The peak at intermediate $\cos \vartheta_{b \bar b}$,
well visible in the case of a
light mass Higgs, corresponds to the contribution of the signal diagram.
The different position of the Higgs peak in the
two-bump shape is a consequence of the different Lorentz boost
acting on the Higgs boson depending on its mass value. The residual bump at
intermediate $\cos \vartheta_{b \bar b}$ for $M_H=140$~GeV comes from
$Z \to b \bar b$ in the $\gamma Z$ contributions.
Figure~5 shows two typical photon energy distributions (for $M_H=65$~GeV
and $140$~GeV, respectively) where the peak at very low energies
is due to dominating soft multiphoton emission.

To summarize, we presented the results of a Monte Carlo event generator for
the four-fermion reaction $e^+ e^- \to \mu^+ \mu^- b \bar b$, including all
the Standard Model diagrams and the effect of initial state radiation in
the leading-log approximation.

\vfill\eject

\leftline{\large \bf Figure Captions}
\vskip 30pt
\noindent
Figure 1. (a) The total cross section of $e^+ e^- \to \mu^+ \mu^- b \bar b$
without (dashed, dash-dotted and closely dotted line)
and with (solid lines) initial state leading-log
 QED corrections as a function of the c.m. energy and for
 different Higgs masses. The sparsely dotted line is the QED corrected total
cross section obtained excluding the Higgs signal.
(b) The ratio $\sigma_s / \sigma_f$
as a function of the Higgs mass in the energy range of LEP2.
$\sigma_s$ is the Higgs signal cross section, $\sigma_f$ the full cross
section including signal and backgrounds. QED corrections are included.

\vskip 8pt\noindent
Figure 2. The $b \bar b$ invariant mass distribution for four different
Higgs masses at $\sqrt s =$~205~GeV (log. scale).
Initial state radiation is included.

\vskip 8pt\noindent
Figure 3. The $b-$quark scattering angle distribution for four different
Higgs masses at $\sqrt s =$~205~GeV. Initial state radiation is included.

\vskip 8pt\noindent
Figure 4. The $b \bar b$ relative angle distribution for four different
Higgs masses at $\sqrt s =$~205~GeV. Initial state radiation is included.

\vskip 8pt\noindent
Figure 5. The photon energy distribution for two Higgs mass values
 at $\sqrt s =$~205~GeV (log. scale).


\begin{thebibliography}{9}

\bibitem{dsc}{See for instance: \\
D.~Schaile, Forsch.~f.~Phys. 42 (1994) 429; \\
G.~Altarelli, {\sl Electroweak Precision Tests: a Status
Report}, CERN Preprint CERN-TH.~7464/94. }

\bibitem{prlcdf}{CDF Collaboration,
F.~Abe et al., Phys.~Rev.~Lett.~73 (1994) 225; Phys.~Rev.~D50 (1994) 2966.}

\bibitem{djou}{For a recent review of the Higgs phenomenology at future
colliders see for instance: A.~Djouadi,
 Report no.~UdeM-GPP-TH.~94-01. }

\bibitem{zal}{V.~Barger et al., Phys.~Rev. D49 (1994) 79 and Proceedings
of the Workshop - Munich, Annecy, Hamburg, DESY Report 93-123C, p.~5,
P.~M.~Zerwas Ed.}

\bibitem{cin}{C.-M.~J.~Chen, Jiunn-Wei Chen and
W.-Y.~P.~Hwang, Phys.~Rev. D50 (1994) 4485.}

\bibitem{russi}{E.~Boos, M.~Sachwitz, H.~J.~Schreiber  and
S.~Shichanin, Z.~Phys. C61 (1994) 675.}

\bibitem{blr}{D.~Bardin, A.~Leike and T.~Riemann, {\sl ``Semi-Analytical
Approach to Higgs Production at LEP 2''}, Preprint DESY 94-097,
CERN-TH.~7305/94, LMU 08/94.}

\bibitem{noiww}{G.~Montagna, O.~Nicrosini, G.~Passarino and
F.~Piccinini, {\sl ``Semi-analytical and Monte Carlo Results
for the Production of Four Fermions in $e^+ e^-$
Collisions''}, CERN Preprint CERN-TH.7497/94, to appear in
Phys.~Lett.~B.}

\bibitem{sf}{E.~A.~Kuraev and V.~S.~Fadin, Sov.~J.~Nucl.~Phys. 41 (1985) 466;\\
G.~Altarelli and G.~Martinelli, in {\it Physics at LEP}, CERN Report 86-02,
J.~Ellis and R.~Peccei eds. (CERN, Geneva 1986) \\
O.~Nicrosini and L.~Trentadue, Phys.~Lett. B196 (1987) 551;
Z.~Phys. C39 (1988) 479. For a review see also \\
O.~Nicrosini and L.~Trentadue, in {\it Radiative Corrections for $e^+ e^-$
Collisions}, J.~H.~K\"uhn ed. (Springer, Berlin, 1989), p.~25; in {\it QED
Structure Functions}, G.~Bonvicini ed., AIP Conf. Proc. No.~201
(AIP, New York, 1990), p.~12. }

\bibitem{schoon}{{\tt SCHOONSCHIP}, a program for symbol handling by
M.~Veltman, see H.~Strubbe, Comput. Phys. Commun. 8 (1974) 1.}

\bibitem{hel} {G.~Passarino, Nucl. Phys. B237 (1984) 249.}

\bibitem{james}{F.~James, Rep. Prog. Phys. 34 (1980) 1145.}

\bibitem{rlux}{F.~James, Comput. Phys. Commun. 79 (1994) 111.}

\bibitem{pdg}{Review of Particles Properties, Phys.~Rev. D50 (1994) 1173.}

\end{thebibliography}
\end{document}